\documentclass[a4paper,11pt]{article}
\usepackage{pos}
\usepackage{siunitx}

\renewcommand{\logo}{\relax}

\title{Towards the first axion search results of the Any Light Particle Search II experiment}

\author*[a]{Li-Wei Wei}
\onbehalf{on behalf of the ALPS collaboration}

\affiliation[a]{Deutsches Elektronen-Synchrotron DESY, \\ Notkestr. 85, 22607 Hamburg, Germany}

\emailAdd{li-wei.wei@desy.de}

\abstract{Any Light Particle Search II (ALPS II) is a dual optical cavity enhanced light-shining-through-a-wall (LSW) experiment at DESY in Hamburg looking for axions and axion-like particles with a target search sensitivity of $g_{a \gamma \gamma}$ down to $2 \times 10^{-11}\,\textrm{GeV}^{-1}$ for masses $m_a \leq 0.1\,\textrm{meV}$. Two 120$\,$m long strings of superconducting dipole magnets have been set up, each providing a magnetic field-length product of $560\,\textrm{T}\cdot\textrm{m}$. A resonant optical cavity with a record-worthy storage time of as high as 7$\,$ms has been constructed to encompass one magnet string. During its initial data-taking phase ALPS~II will be operated with a simplified optical configuration that facilitates the characterization of the experiment. The first ALPS~II science run took place in May/June 2023 and achieved an estimated search sensitivity of $g_{a \gamma \gamma}$ of around $6 \times 10^{-10}\,\textrm{GeV}^{-1}$ for $m_a \leq 0.1\,\textrm{meV}$. Data analysis and further data runs are under way. Final results on axion-photon coupling from the initial science runs of ALPS\,II are expected in early 2024.}

\FullConference{
28.08 -- 01.09.2023\\
University of Vienna\\}

\begin{document}
\maketitle

\section{Introduction}
The axion is a highly-motivated hypothetical particle beyond the Standard Model of particle physics. It arises as a pseudo-Nambu-Goldstone boson from the spontaneous breaking of the Peccei-Quinn symmetry proposed to address the Strong CP problem in quantum chromodynamics (QCD) \cite{Peccei+Quinn_1977}. The axion is also a leading candidate particle of Dark Matter. The detection of the axion would therefore solve two puzzles coming from physics of very different scales \cite{Kim+Carosi_2010}.

Experimental attempts on the detection of the axion most often are based on its interaction with photons. In the presence of a magnetic field, an axion can convert into a photon and vice versa, via the so-called Sikivie and Primakoff effects \cite{Primakoff_1951}. The Any Light Particle Search II (ALPS~II) experiment \cite{ALPS_II_TDR_2013} at DESY Hamburg is also devised along these effects.

To address the Strong CP problem, the axion-photon field coupling strength, denoted as $g_{a \gamma \gamma}$ in the interaction Lagrangian $\mathcal{L} = g_{a \gamma \gamma}\,a\,\vec{E} \cdot \vec{B}$, where $a$ is the axion field, and $\vec{E}$ and $\vec{B}$ are the electric and magnetic fields, respectively, must observe a relation with respect to the axion mass $m_a$. The generic properties of the axion can also be used to source a family of axion-like particles (ALPs) whose mass-coupling relation differs from that of the QCD axion, but instead satisfies certain astrophysical observations. The ALPS~II experiment intends to seek exactly these ALPs. 

The canonical benchmark coupling coefficients of the QCD axion is derived for its coupling to fermions (Dine-Fischler-Srednicki-Zhitnitsk, DFSZ model \cite{DFSZ}) and quarks beyond the standard model (Kim-Shifman-Vainshtein-Zakharov, KSVZ model \cite{KSVZ}). Recently, an alternative QCD axion model has been developed on the basis of the coupling of the Peccei-Quinn scalar field to magnetically charged fermions at high energies, and yields QCD axion coupling coefficients that differ drastically from the above-mentioned canonical models \cite{Sokolov_Ringwald_2021}. The ALPS~II experiment is equipped with the sensitivity to also probe such alternative solutions for the QCD axion.

\begin{figure}[b]
    \centering\includegraphics[scale=0.4]{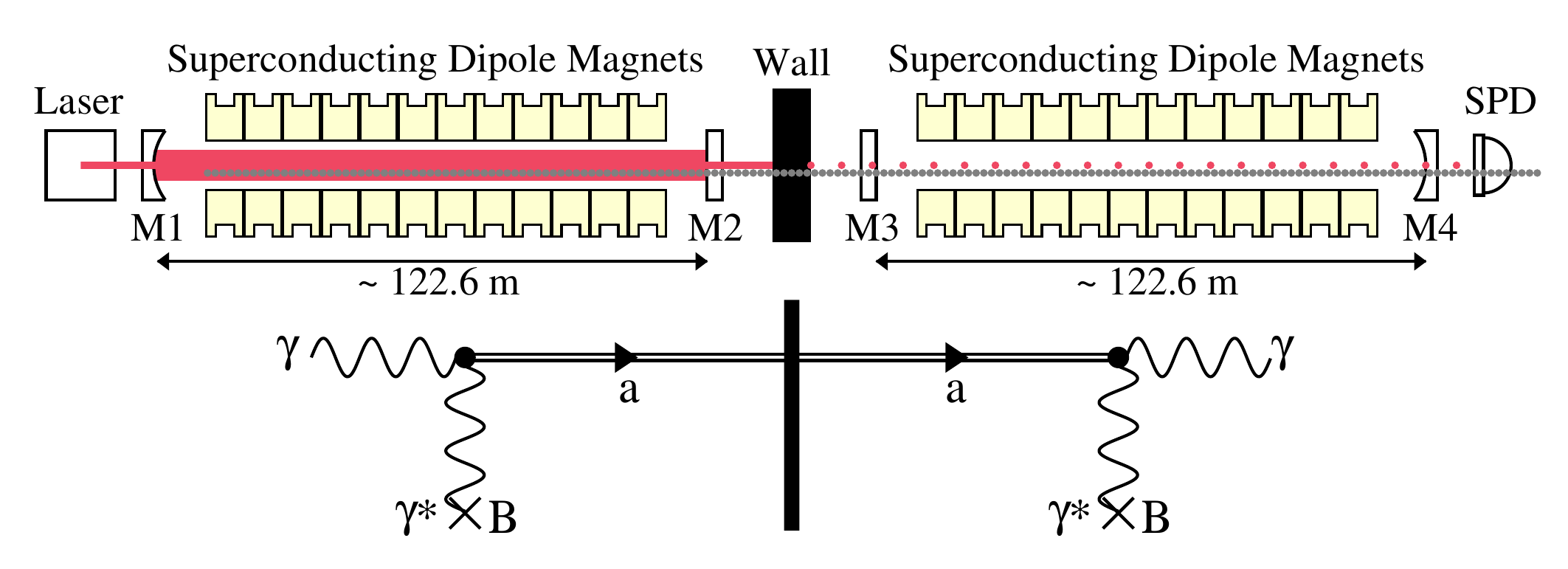}
    \caption{Schematic of the ALPS\,II experiment and the Feynman diagram illustrating the light-shining-through-a-wall process $\gamma \rightarrow a \rightarrow \gamma$ mediated by an axion-like particle a in the presence of a magnetic field B. M: Mirror; M1 and M2 form the (axion-) Production Cavity, and M3 and M4 form the (photon-) Regeneration Cavity; SPD: single-photon detector; dotted gray: axion-like particles; dotted red: regenerated photons.}
    \label{fig:alps_ii_schmatic}
\end{figure}

\section{The ALPS~II Experiment}

ALPS\,II is a dual optical cavity enhanced light-shining-through-a-wall (LSW) \cite{LSW} experiment located at DESY Hamburg. Figure \ref{fig:alps_ii_schmatic} shows the schematic of ALPS\,II. ALPS\,II aims at improving the search sensitivity by 3 orders of magnitude in terms of $g_{a \gamma \gamma}$ in comparison to previous LSW experiments \cite{ALPS_2010}. The design sensitivity of $g_{a \gamma \gamma}$ of $\SI{2E-11}{\per\giga\electronvolt}$ of ALPS\,II delves into uncharted parameter space motivated by astrophysical observations \cite{ALPS_II_hints} in a pure laboratory setting that is free from complex model assumptions.

At the target search sensitivity, the projected regenerated photon rate in ALPS\,II is:
\begin{equation}
    \dot{N}_{\gamma} \approx \SI{4e-5}{\hertz} \cdot \left( \frac{P_\textrm{laser}}{\SI{40}{\watt}} \right) \cdot \left( \frac{\lambda_\textrm{laser}}{\SI{1064}{\nano\meter}} \right) \cdot \left( \frac{\beta_\textrm{PC}}{5000} \right) \cdot \left(\frac{B\cdot L}{\SI{560}{\tesla\cdot\meter}} \right)^4 \cdot \left( \frac{g_{a \gamma \gamma}}{\SI{2E-11}{\per\giga\electronvolt}} \right)^4 \cdot \left( \frac{\beta_\textrm{RC}}{40000} \right),
\nonumber
\end{equation}
where $P_\textrm{Laser}$ is the laser power, $\lambda_\textrm{Laser}$ is the laser wavelength, $\beta_\textrm{PC}$ is the power build-up of the PC, $B\cdot L$ is the magnetic field-length product, and $\beta_\textrm{RC}$ is the power build-up of the RC.

The ALPS\,II high-power laser source is a master oscillator power amplifier (MOPA) system at \SI{1064}{\nano\meter} wavelength \cite{ALPS_II_HPL} that has a power output of about \SI{60}{\watt}. After spatial-mode filtering and the injection optics, around \SI{40}{\watt} of laser power is currently injected into the ALPS\,II experiment.

ALPS\,II reuses superconducting dipole magnets from the former HERA accelerator to construct its two magnet strings. Each string consists of 12 magnets. The magnets are straightened to increase the aperture from $\approx\SI{37}{\milli\meter}$ to $\approx \SI{50}{\milli\meter}$. Each magnet has a physical length of $\approx \SI{10}{\meter}$, and provides a magnetic field of $\SI{5.3}{\tesla}$ over $\approx \SI{8.8}{\meter}$ \cite{ALPS_II_magnets}. The magnet strings were first powered on to a current level of $\SI{5700}{\ampere}$ for a successful test in March 2022.

The power build-up of the optical cavities is given by $\beta = 4 \cdot T_\textrm{input} / \rho^2$, where $T_\textrm{input}$ is the input mirror transmissivity and $\rho$ is the round-trip loss of the cavity. The product of $P_\textrm{laser}$ and $\beta_\textrm{PC}$ gives the intracavity power of the PC. The laser intensity that can be withstood by the cavity mirrors sets a limit on the product of $P_\textrm{laser}$ and $\beta_\textrm{PC}$. The power build-up of RC is limited by its round-trip loss, which in turn depends on the quality of the coated cavity mirrors. ALPS\,II has acquired cavity mirrors that project an expected $\beta_\textrm{RC}$ of $\approx \num{16000}$ for the initial science run \cite{ALPS_II_Optics_PDU}. Acquisition of cavity mirrors of better quality is also planned.

The projected photon rate of $\dot{N}_\gamma \approx \SI{4e-5}{\hertz}$ is equivalent to $\approx 3.5$ photons per day, and requires high-sensitivity single-photon counters. ALPS\,II anticipates two detectors, the transition edge sensor (TES), which is a super sensitive calorimeter operated at cryogenic temperatures, and the heterodyne detector, which is based on the coherence properties of electromagnetic fields. The TES system developed for ALPS\,II has demonstrated an intrinsic dark count rate as low as \SI{6.9e-6}{\hertz} \cite{Shah_2022}. The heterodyne detection scheme for ALPS\,II has also been shown to have a dark count rate of $\approx 10^{-5}\,\si{\hertz}$ after an integration time of $10^6$\,s \cite{Bush_2019}. Each of the two detection schemes requires a distinct optics setup. The two detection schemes therefore cannot be operated in parallel.

\section{Initial Science Run}

\begin{figure}[t]
    \centering\includegraphics[scale=0.35]{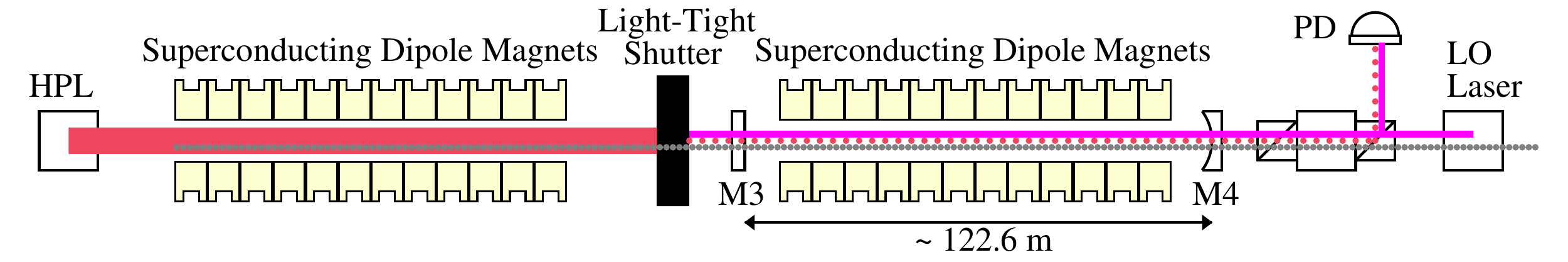}
    \caption{Schematic of the ALPS\,II initial science run with the heterodyne detection scheme. HPL: high-power laser; M3 and M4 form the (photon-) Regeneration Cavity. The local oscillator (LO) laser interferes with the regenerated photon signal to generate a radio-frequency signal detected by the photodiode (PD).}
    \label{fig:alps_ii_isr_schmatic}
\end{figure}

The installation of the ALPS\,II experimental setup with the heterodyne detection scheme was completed in September 2022. To facilitate stray light hunting and mitigation, ALPS\,II is carrying out its initial science run without the PC in place; the laser power impinging on the wall is increased by a factor of around 40. The wall is in practice a light-tight shutter. By operating the experiment with the magnets off and the shutter closed, we measure the background of the heterodyne detection scheme in situ. The schematic of the ALPS\,II initial science run is shown in Figure \ref{fig:alps_ii_isr_schmatic}.

We have so far measured a cavity storage time of up to \SI{7.04}{\milli\second} for the RC. To the best of our knowledge, this corresponds to the longest storage time of a two-mirror optical cavity reported to date. Nonetheless, the excess cavity round-trip loss is higher than anticipated, and the corresponding $\beta_\textrm{RC}$ is \num{7705}, which falls short of the anticipated value of \num{16000}. We notice the dependence of the cavity round-trip loss with respect to the laser beam spot position on the cavity mirrors, and are looking into a more comprehensive study of the effect to potentially identify locations that result in higher $\beta_\textrm{RC}$.

The initial science run of ALPS\,II took place in May/June 2023. The polarization of the high-power laser was set to be perpendicular to that of the magnet field. The experiment in such configuration is insensitive to pseudo-scalar particles such as the axion and therefore enables a measurement of the background (as well as scalar particles, whose coupling to photons is nevertheless much more stringently bounded). We have accumulated around $\SI{150000}{\second}$ of science data. First analysis on the data has identified stray light as the main limit in the sensitivity of the initial science run of the ALPS\,II experiment. The estimated sensitivity reach in $g_{a \gamma \gamma}$ based on the first data is around $\SI{6e-10}{\per\giga\electronvolt}$ for $m_a \leq \SI{0.1}{\milli\electronvolt}$. ALPS\,II will continue to take data in its current configuration until early 2024 to accumulate $10^6\,\si{\second}$ worth of data. 

In the heterodyne detection scheme of ALPS\,II the weak axion-photon signal is interfered with a strong electromagnetic field with controlled frequency offset of several tens of $\si{\mega\hertz}$ between the HPL and the LO laser (Figure \ref{fig:alps_ii_isr_schmatic}). For the signal to remain coherent for the duration of heterodyne detection, the frequency offset needs to be stable to better than $\SI{3}{\micro\hertz}$ over the integration time of $10^6\,\si{\second}$. The coherence time of stray light on the other hand is expected to be shorter than that of the axion-photon signal. The initial science run data suggests a destructive interference of the stray light at the $10^6\,\si{\second}$ timescale such that it stops being the dominating background of the experiment.







\section{Summary and Outlook}
After years of preparation, ALPS\,II started data-taking in May 2023 and has achieved an estimated sensitivity reach that is close to two orders of magnitude better than its predecessors so far. ALPS\,II is currently operating without the PC in place to facilitate the identification and mitigation of stray light. First data suggest that the impact of stray light might be mitigated by extending the integration time to $10^6\,\si{\second}$. ALPS\,II is currently taking data in the search mode for scalar particles. This will be followed by the search for pseudo-scalar particles. Final results on axion-photon coupling from the initial science runs of ALPS\,II are expected in early 2024.

The PC will be installed during the course of 2024 after the initial science runs, which will improve the sensitivity reach of ALPS\,II via its power build-up as well as the reduction of laser power impinging on the wall, which we believe is the main source of stray light. Upgraded mirrors for the RC are being planned for late 2024 / early 2025 to improve the sensitivity of ALPS\,II towards its final design goal of $2 \times 10^{-11}\,\textrm{GeV}^{-1}$ in terms of $g_{a \gamma \gamma}$ for masses $m_a \leq 0.1\,\textrm{meV}$.

\bigskip

\noindent
\textbf{Acknowledgments.}
We are grateful to numerous enthusiastic engineers, technicians, administrators and other helpful colleagues at DESY and elsewhere who contributed to get ALPS II going. We acknowledge support by the Deutsche Forschungsgemeinschaft (Germany), European Research Council (grant No. 948689), Max-Planck Society (Germany), National Science Foundation (USA, NSF PHY-2309918), Heising-Simons Foundation (USA, 2020-1841), Helmholtz Association (Germany), Science and Technology Facilities Council (UK, ST/T006331/1) and the Volkswagen Stiftung (Germany).

\end{document}